*Article*

# Relaxation Behavior by Time-Salt and Time-Temperature Superpositions of Polyelectrolyte Complexes from Coacervate to Precipitate[†]

**Samim Ali[⊥] and Vivek M. Prabhu ***

Material Measurement Laboratory, National Institute of Standards and Technology, 100 Bureau Drive, Gaithersburg, MD 20899, USA;
[⊥] Guest Researcher: samim.ali@nist.gov
* Correspondence: vprabhu@nist.gov; Tel.: +1-301-975-3657
[†] Official contribution of the National Institute of Standards and Technology; not subject to copyright in the United States.

**Abstract:** Complexation between anionic and cationic polyelectrolytes results in solid-like precipitates or liquid-like coacervate depending on the added salt in the aqueous medium. However, the boundary between these polymer-rich phases is quite broad and the associated changes in the polymer relaxation in the complexes across the transition regime are poorly understood. In this work, the relaxation dynamics of complexes across this transition is probed over a wide timescale by measuring viscoelastic spectra and zero-shear viscosities at varying temperatures and salt concentrations for two different salt types. We find that the complexes exhibit time-temperature superposition (TTS) at all salt concentrations, while the range of overlapped-frequencies for time-temperature-salt superposition (TTSS) strongly depends on the salt concentration ($C_s$) and gradually shifts to higher frequencies as $C_s$ is decreased. The sticky-Rouse model describes the relaxation behavior at all $C_s$. However, collective relaxation of polyelectrolyte complexes gradually approaches a rubbery regime and eventually exhibits a gel-like response as $C_s$ is decreased and limits the validity of TTSS.

## 1. Introduction

A mixture of anionic and cationic polyelectrolytes in an aqueous medium undergoes complexation resulting in a polymer-rich phase coexisting with a polymer-poor liquid phase. The dense phase can be liquid-like coacervates, or solid-like precipitates depending on the salt concentration, charge-stoichiometric ratio, pH, chain lengths and hydrophobicity of the polymers [1–5]. It is argued that the complexation is driven by a combined effect of entropic and enthalpic interactions between polymer chains [3,6]. The complexation behavior of oppositely-charged species occurs across a wide-variety of charged macromolecules including synthetic and bioderived polymers [7,8], copolymers [9], proteins [10–13], and polymer/surfactant mixtures [14,15]. In general, a coacervate is a viscoelastic liquid [16–18] that exhibits very low interfacial tension at the coacervate/dilute phase boundary [19–21]. A coacervate can also exhibit strong adhesion [22,23]. On the other hand, a complex precipitate, a water-poor phase, behaves more like a viscoelastic solid resembling a physically crosslinked gel. Owing to these tunable properties, polyelectrolyte complexation has garnered interest for various applications including surgical wet adhesives [24], tissue engineering [25], protein purification [26], food processing, water purification [27], coating and packaging [28] by changing the complexation state from liquid-like to solid-like, in situ. For such applications, it is necessary to understand how structural and dynamical properties evolve during the transition of complexes prepared at various physicochemical conditions.

The electrostatic interactions between polyelectrolyte chains and their relaxation mechanism within the complex phase are of fundamental interest [29]. Such interactions were modeled by thermoreversible bond formation via ion-pairing between anionic and cationic polyelectrolytes segments [30–33]. At low salt concentration, the ion-pair interaction is strong and results in solid-like precipitate. Addition of salt reduces the number of ion-pairs and weakens the ion-pair strength resulting in sticky, but labile connections between oppositely-charged segments that results in a liquid-like coacervate [16,34]. Attempts to characterize the transition between these two distinct phases were made by comparing the magnitudes of elastic $G'$ and viscous $G''$ moduli measured at a fixed frequency [32] or by comparing viscoelastic data measured over a frequency range of 1 rad/s to 100 rad/s [35]. The boundary between the liquid-like coacervate phase and solid-like precipitate phase is broad and was identified as a continuum by Wang and Schlenoff [32]. The relaxation mechanism of polymer chains across this continuum thus expected to exhibit gradual evolution. The complete understanding of the changes in the relaxation process of polymer chains across the precipitate-coacervate continuum therefore requires measurement of dynamics over a wide range of timescales.

In the present work, the changes in the dynamics of model polyelectrolyte complexes over a wide timescale are measured across the coacervate–precipitate continuum driven by salt concentration. For this purpose, viscoelastic spectra and zero-shear viscosities are measured at varying temperatures and salt concentrations. All the measurements and subsequent analysis are repeated for complexes of sodium poly(styrene sulfonate) (NaPSS) and poly(diallyl dimethyl ammonium chloride) (PDADMAC) with added NaCl, as well as potassium poly(styrene sulfonate) (KPSS) and poly(diallyl dimethyl ammonium bromide) (PDADMAB) with added KBr. The complexes prepared with two different salt types with matching counterions allows generalization of the observed behavior. The validity of time-temperature superposition (TTS) and time-temperature-salt superposition (TTSS) will be examined and the relaxation times further analyzed by the main results of the sticky-Rouse model [36].

## 2. Results and Discussion

A mixture of anionic and cationic linear polyelectrolytes in aqueous medium exhibits different phases depending on the concentrations and sizes of salt ions. At low salt concentrations of NaCl, aqueous solutions of NaPSS and PDADMAC mixed at 1:1 charge-stochiometric ratio exhibits phase separation into polymer-poor supernatant and polymer-rich precipitates that are stiff and appear turbid as shown in Figure 1a,b for two representative salt concentrations for $C_{NaCl} \leq 2$ M (M has been used to represent the International System of Units (SI) unit mol/L to conform to the requirement of the Journal). The polymer-rich phase gradually becomes softer and less turbid with increasing amount of salt as shown in Figure 1c,d for $C_{NaCl}$ = 2.4 and 2.6 M, respectively. As the salt concentration is increased further, the polymer-rich phase becomes more liquid-like and transparent (Figure 1e–g for $C_{NaCl} \geq 3$ M). A uniform solution (no complexation) is not observed even for $C_{NaCl} >$ 4.6 M for the molecular mass of polymers used in the current experiments.

The replacement of NaCl with KBr salt has a noticeable effect on the complexation behavior due to change in the sizes of the hydration shells of dissociated salt ions [37]. In this case, we observe a similar trend, however, the phase boundaries shift to lower salt concentrations with the appearance of stiff and turbid precipitate for $C_{KBr}$ < 0.8 M (Figure 1h). Increasing $C_{KBr}$ leads to softening of the dense phase as shown in Figure 1i for $C_{KBr}$ < 0.8 M. When the salt concentration is increased further, the polymer-rich phase becomes gradually more liquid-like (Figure 1j–l for 1.4 M $\leq C_{KBr} \leq$ 1.8 M). A one-phase solution appears for $C_{KBr}$ > 1.85 M (Figure 1m), the regime where no complexation occurs between the oppositely-charged polyelectrolytes.

The visual observations presented in Figure 1a–g and Figure 1h–l for two different 1:1 salts clearly show that the boundaries between liquid-like coacervate and solid-like precipitate phases are quite broad. Moreover, it is not possible to define any distinct boundary of the rheological properties between these phases based on the visual appearance. The transition between these two distinct complex phases requires a systematic study of the polymer dynamics over the entire salt concentration range, or precipitate-coacervate continuum [32].

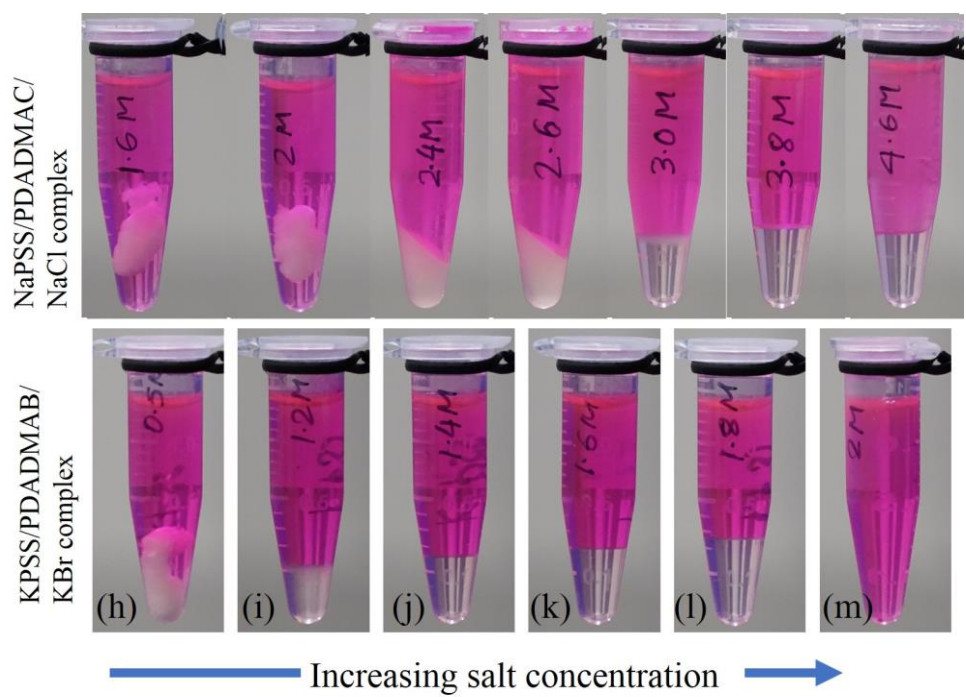

**Figure 1.** Photographs of complexes obtained after centrifugation for polyelectrolyte combinations of (**a**–**g**) NaPSS/PDADMAC for several concentrations of NaCl, (**h**–**l**) KPSS/PDADMAB at several concentrations of KBr with (**m**) uniform solution at 2 M KBr. Rhodamine-B dye was added to the supernatant to show the interface between supernatant at the top and complexes at the bottom in the microcuvettes.

Zero-shear viscosity and dynamic rheological measurements were performed to understand the evolution of the interactions between oppositely charged polyelectrolytes in the precipitate-coacervate continuum. This will provide insight into the underlying polyelectrolyte complex relaxation dynamics over a wide time and associated length scales. The linear-viscoelastic spectra of complexes at varying salt concentrations and temperatures ($T$) are measured by applying oscillatory strains in the frequency ($\omega$) range 0.01 to 200 rad/s. Figure 2a shows typical viscoelastic spectra at varying temperatures for $C_{NaCl}$ = 4.6 M. This sample exhibits behavior typical for a viscoelastic-liquid with $G'' > G'$ over a wide frequency range. At $T$ = 10 °C, a crossover between $G'$ and $G''$ can be seen at $\omega$ = 60 rad/s. However, the crossover point disappears at temperatures $T$ > 20 °C. The viscoelastic spectra acquired in the range 10 to 60 °C are found to obey TTS and all the frequency-dependent viscoelastic data are superposed onto a master-curve as shown in Figure 2b. The superposition provides more than six decades in relaxation behavior to probe the chain dynamics over a wide range of timescales. The inset of Figure 2b shows that the temperature dependence of shift factors $a_T$ that can be described by the Williams-Landel-Ferry (WLF) relation: $\ln(a_T) = -C_1(T - T_r)/(C_2 + T - T_r)$, where $C_1$ = 7.3, $C_2$ = 45 are material-dependent parameters and $T_r$ = 20 °C is the reference temperature. The inset of Figure 2b also shows that the shift factors $a_T$ obtained using the superposition algorithm are nearly equal to $a'_T = \eta'_0(T)/\eta_{r0}$, where $\eta'_0(T)$ and $\eta_{r0}$ are zero-shear viscosities at temperatures $T$ and $T_r$, respectively. The applicability of TTS suggests that the shape of the relaxation spectra of the complexes is independent of temperature.

Such a self-similarity implies that the variation of temperature influences the relaxation rate without altering the relaxation mechanism itself within the temperature range of 10 to 60 °C.

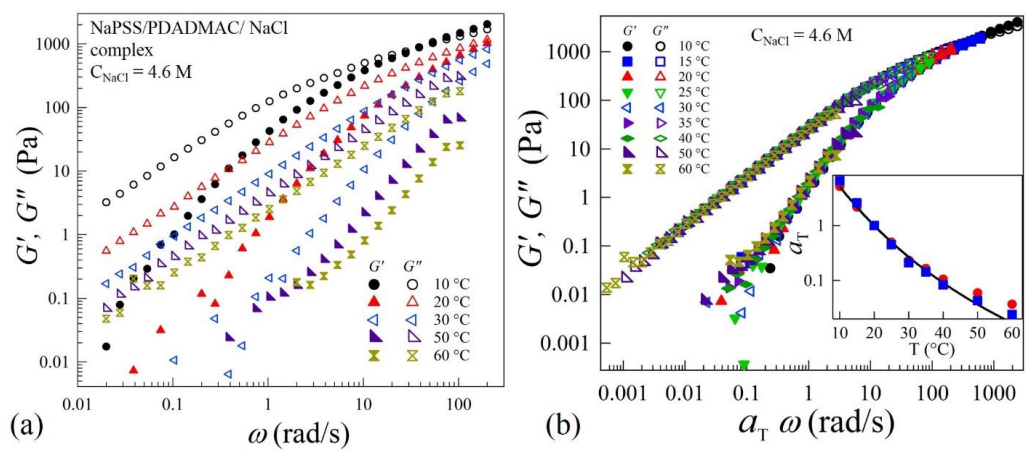

**Figure 2.** (**a**) Variations of elastic $G'$ and viscous $G''$ moduli as functions of applied frequency $\omega$ at different temperatures in linear viscoelastic regime of coacervates prepared at $C_{NaCl}$ = 4.6 M, (**b**) time-temperature superposition (TTS) at a reference temperature $T_r$ = 20 °C of the data shown in Figure 2a. The inset of Figure 2b shows the dependence of shift factors $a_T$ (■) and $a'_T$ (●) on temperature $T$. The solid line is a fit to the Williams–Landel–Ferry (WLF) equation: $\ln(a_T) = -C_1(T - T_r)/(C_2 + T - T_r)$, where $C_1$ = 7.3, $C_2$ = 45 and $T_r$ = 20 °C.

We next show that TTS applies for all complexes prepared at different salt concentrations corresponding to the different phases along the precipitate-coacervate continuum. Figure 3a,c show selected (for clarity) data of TTS master curves obtained at a reference temperature $T_r$ = 20 °C for complexes prepared with NaCl and KBr, respectively. The shift factors $a_T$ follow the WLF equation as presented in Figure 3b,d. It can be seen in Figure 3a,c that the complexes prepared at $C_{NaCl}$ ≥ 3.8 M and $C_{KBr}$ ≥ 1.5 M exhibit near-terminal flows where $G'' > G'$ and $G' \sim \omega^{1.7}$ and $G'' \sim \omega^1$ at low frequencies. The complexes therefore are liquid-like coacervates. The deviation of the power-law growth of $G'$ from a square-law ($G' \sim \omega^2$) at low frequencies and the monotonic increase of $G'$ and $G''$ at higher frequencies after the crossover may indicate multiple relaxation times in the coacervate systems. For these coacervates, the difference $\delta G = G'' - G'$ decreases with increasing $\omega$ and finally a crossover between $G'$ and $G''$ at $\omega = \omega_g$ (shown by an arrow in Figure 3a) can be observed at higher frequencies for samples with NaCl, indicating solid-like response at short timescales. This crossover can be associated with the relaxations of chain segments between two consecutive ion-pairs of oppositely-charged sites on the polyelectrolytes backbones [38]. The position of $\omega_g$ shifts to lower frequencies as salt concentration is decreased to $C_{NaCl}$ = 2.8 M. Interestingly, a similar crossover is not observed in the KBr samples for $C_{KBr}$ >1.5 in the frequency range measured here (Figure 3c). This suggests a more liquid-like coacervate in the presence of the KBr salts. However, the viscoelastic spectra shift systematically towards higher moduli magnitudes as the salt concentration decreases in both systems. It should be noted that the viscoelastic response of samples with $C_{KBr}$ ≥ 1.7 M were not measured at $\omega$ < 1 rad/s due to the very low torque response that is below the noise level of the rheometer transducer.

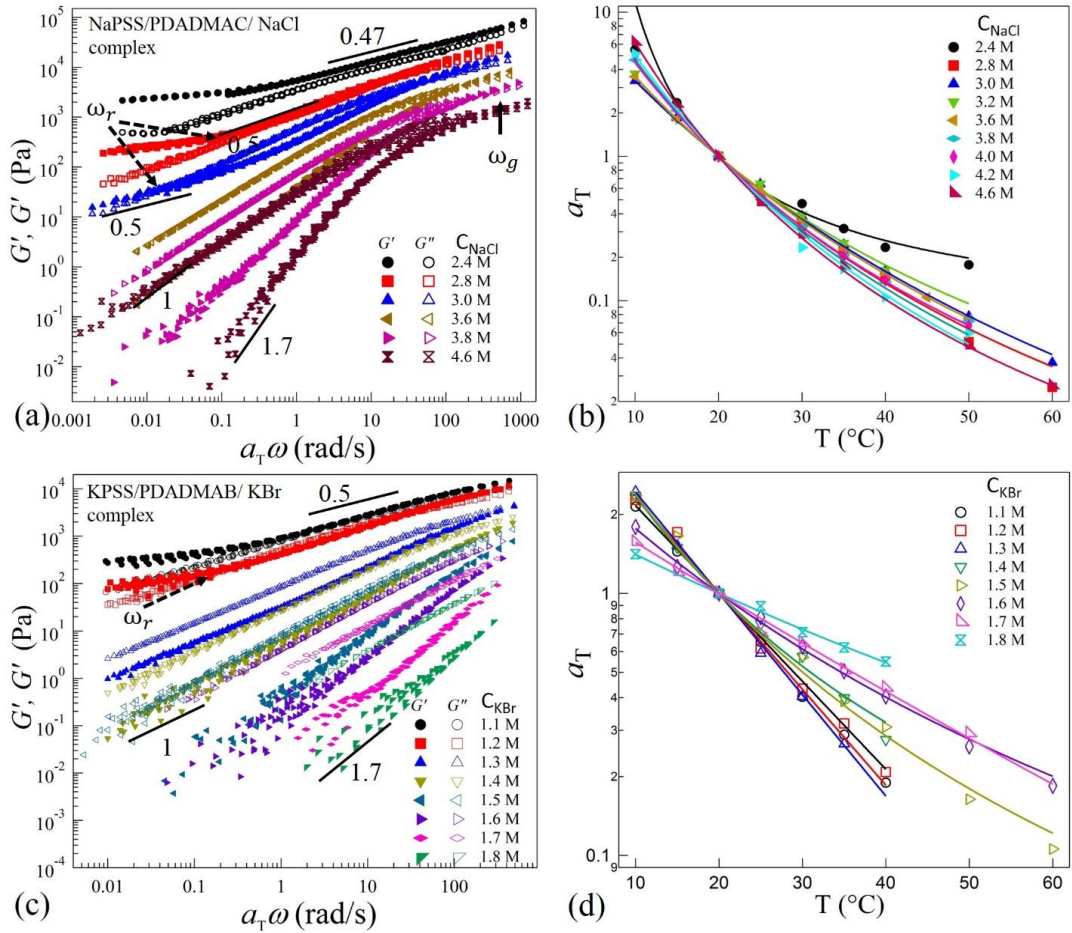

**Figure 3.** Plots of master curves and corresponding shift factors $a_T$ obtained after performing TTS of frequency-dependent viscoelastic moduli $G'$ and $G''$ with $T_r$ = 20 °C for complexes prepared at varying concentrations of (**a**,**b**) NaCl. Data for $C_{NaCl}$ = 2.8 M and 2.4 M are shifted vertically upward by a multiplicative factor 2 for clarity, and (**c**,**d**) KBr. The solid lines in Figure 3b,d are fit to the WLF equation.

When the salt concentration is decreased to $C_{NaCl}$ = 3.6 M (Figure 3a) and $C_{KBr}$ = 1.3 M (Figure 3c), both the moduli follow $G'$, $G'' \sim \omega^1$ with $G'' > G'$ over three decades of frequencies, indicating a departure from the terminal-flow regime. Interestingly, as the salt concentration is decreased further, the complexes with $C_{NaCl} \leq 3.0$ M and $C_{KBr} \leq 1.2$ M exhibit second crossover-point at $\omega = \omega_r$ (shown by dotted arrows in Figure 3a,c) in the low frequency regime. This crossover-point appears to be associated with the transition to the rubbery relaxation regime as the complexes exhibit a crossover to $G' > G''$ as the $\omega$ is decreased. Moreover, both the moduli approach plateau values at longer timescales for $\omega < \omega_r$. We also observe that the position of crossover-point $\omega_r$ shifts to higher frequencies as salt concentration is decreased

As the salt concentration is decreased further, the crossover points $\omega_r$ and $\omega_g$ approach each other as seen for $C_{NaCl}$ = 2.8 M and $C_{KBr}$ = 1.1 M in Figure 3a,b, respectively. Moreover, these complexes appear to exhibit a behavior similar to the Winter-Chambon criterion: $G' = G'' \sim \omega^{0.5}$ over a wide frequency range [39]. However, at low frequency, a deviation in the scaling as a rubbery-plateau appears. Therefore, these coacervates are not in a critical-gel state. Similarly, when the salt concentration is decreased to $C_{NaCl}$ = 2.4 M, the elastic behavior dominates ($G' > G''$) in the transition zone; however, both moduli exhibit $\approx \omega^{0.47}$ dependence over three decades. This is indicative of increased physical crosslinking that leads to stiffer complexes. Rheological measurements were not performed for samples with $C_{NaCl} < 2.4$ M (Figure 1a,b) and $C_{KBr} < 1.1$ M (Figure 1h) as these samples are too stiff and could not be loaded into the sample cell appropriately.

Moreover, measurements on these solid-like samples may lead to wall-slip conditions that require an alternate analysis [40,41]. Nonetheless, the measured values of salt concentrations of NaCl and KBr in the current work provide substantial ranges to probe the evolution of relaxation dynamics in the precipitate-coacervate continuum.

To identify the structural similarities of the complexes with changing salt concentrations, we apply TTSS to the rheological data. We find that the frequency-dependent viscoelastic moduli $G'$ and $G''$ acquired at various temperatures and salt concentrations can only be superposed onto a master curve over the entire frequency range for $C_{NaCl} \geq 3.8$ M and $C_{KBr} \geq 1.5$ M as shown in Figure 4a,b for reduced salt concentrations $C^r_{NaCl} = 4.6$ M and $C^r_{KBr} = 1.8$ M, respectively. On the other hand, the linear viscoelastic moduli for $C_{NaCl} \leq 3.6$ M and $C_{KBr} \leq 1.4$ exhibit superposition within a limited frequency range around the crossover-point at $\omega = \omega_g$ in the high frequency regime, with the width of the overlapping frequency-range shortening as the salt concentration decreases. It should be noted that such a deviation from the TTSS at low frequency occurs even for the complexes exhibiting liquid-like coacervate ($G' < G''$) in the intermediate salt concentration range 3.2 M < $C_{NaCl}$ < 3.8 M and 1.3 M < $C_{KBr}$ < 1.6 M. The insets of Figure 4a,b show the salt concentration dependence of the shift factors $a_s$ and $b_s$ associated with the horizontal and vertical shifts, respectively. We find that $b_s$ exhibits a weak dependence on the salt concentration, while the magnitude of $a_s$ sharply increases with the decreasing salt concentration. The variation of the shift factor $a_s$ provides important information about the chain relaxation mechanism as $a_s$ is directly proportional to the relaxation time $\tau_0$ associated with the formation and breaking of the ion pairs between anionic and cationic polyelectrolytes.

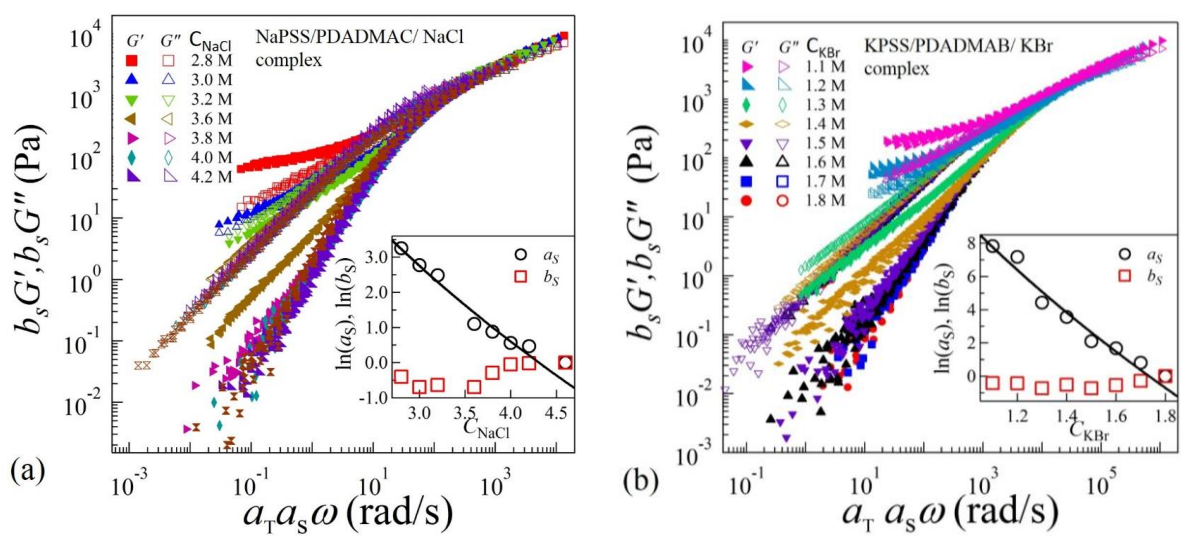

**Figure 4.** Time-temperature-salt superposition of frequency sweep data acquired in complex coacervates in the presence of (**a**) NaCl at reduced salt concentration $C^r_{NaCl} = 4.6$ M and (**b**) KBr at reduced salt concentration $C^r_{KBr} = 1.8$ M. The insets show plot of the shift factors $a_s$ and $b_s$ as functions of salt concentrations. The solid lines are fit to Equation (3).

The relaxation mechanism within the precipitate-coacervate continuum may be modeled by sticky-Rouse dynamics [36]. In this model, segments of the oppositely-charged polymers form ion-ion pairs (sticky points) with a lifetime that depends on the concentration of salt and counterions and polymers in the solution. The polymer chains in a complex reorganize via reversible breaking and reformation of these ion pairs. The relaxation behavior of these sticky points therefore controls the chain dynamics. The validity of the sticky-Rouse model in the coacervate regime was established for weak polyelectrolytes [16,42]. In the present work, we want to verify whether this model still explains the relaxation behavior in the precipitate-coacervate continuum. According to this model, the average time two oppositely-charged sites remains associated can be approximated as [31]

$$\tau_0 = \left(\frac{1}{\omega_0}\right)\exp(E_a(C_{salt})/k_B T), \qquad (1)$$

where $\omega_0$ is relaxation rate in the absence of any ionic interactions, $k_B$ is the Boltzmann constant, $T$ is the temperature and $E_a$ is the activation energy associated with the ion pairs. The magnitude $E_a$ can be estimated as the difference between correlation energy $E_{corr}$ of separated charged groups in salt solution in an unbound state and the Coulomb energy $E_{Coul}$ of ion pairs in the bound state. Using the Debye–Hückel approximation for $E_{corr}$, Spruijt et al. derived an expression for $E_a$ [42],

$$\frac{E_a(C_{salt})}{k_B T} = -P\sqrt{C_{salt}} + Q. \qquad (2)$$

$P = (32 \times 10^3 \pi N_A l_B^3)^{1/2}$ is associated with the rearrangement of single ionic bonds, $Q = 2l_B/d$, $l_B$ is the Bjerrum length and $d$ is the effective distance of ionic bonds and $N_A$ is Avogadro's number. Combining Equations (1) and (2), we get

$$\ln\tau_0 = -P\sqrt{C_{salt}} + Q - \ln\omega_0. \qquad (3)$$

The magnitude of $\omega_0$ depends on the molecular mass and concentration of polymers in solution. In the semidilute entanglement regime for polymers in good solvent, we assume $\omega_0 \propto \varphi_p^{-7/3}$, where $\varphi_p$ is polymer concentration [43].

In the sticky-Rouse model, $\tau_0$ is associated with the segmental relaxation time between two sticky points. The magnitude of horizontal shift factors $a_s$ associated with TTSS should therefore follow $a_s \propto \tau_0$. Interestingly, we find that the variation of $a_s$ shown in the insets of Figure 4a,b can be fit to Equation (3) with $P = 6.8 \pm 0.6$ and $23.8 \pm 2.5$ for complexes with NaCl and KBr salts, respectively. The uncertainties (±) are estimated by one standard deviation of the linear regression fit parameters. In these fits, the polymer concentration was experimentally determined as $\varphi_p = 880 C_{NaCl}^{-0.5}$ and $\varphi_p = 569 C_{KBr}^{-1.3}$, where $\varphi_p$ is a function of salt concentration as shown in the insets (c) and (e) of Figure 5a,b, respectively. Interestingly, the magnitudes of $P$ do not differ much if the concentration-dependence of $\omega_0$ in Equation (3) is ignored, as was assumed previously [16].

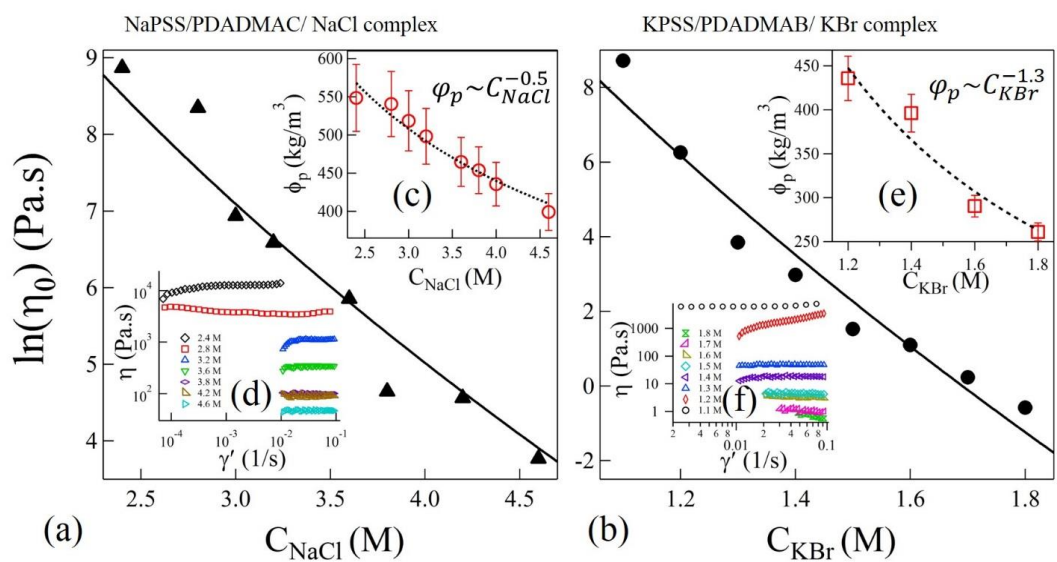

**Figure 5.** Variation of zero-shear viscosities $\eta_0$ as functions of (**a**) $C_{NaCl}$ and (**b**) $C_{KBr}$. Solid lines are fit to Equation (4). The insets (**c**,**e**) in Figure 5a,b show the variation of polymer concentrations in complexes and power-law fits (dotted lines) as functions of $C_{NaCl}$ and $C_{KBr}$, respectively. Shear viscosities (η) as functions of shear-rate (γ′) at varying concentrations NaCl and KBr are shown in the insets of (**d**,**f**), respectively.

The validity of the modeling of ion-pair relaxations in the coacervate-precipitation continuum zone can be further verified by using the expression of zero-shear viscosity predicted by the sticky-Rouse model [36],

$$\eta_0 = \frac{k_B T \varphi_p \tau_R}{b^3 N} = N^{\beta_1} \varphi_p(N, C_{salt})^{\beta_2} \exp(-P\sqrt{C_{salt}} + Q), \quad (4)$$

where $b$ is the Kuhn length and $\varphi_p/b^3N$ is the number density of chains. The shear viscosities (η) as functions of shear rate (γ′) at 20 °C and at selected values of $C_{NaCl}$ and $C_{KBr}$ are shown in the insets (d) and (f) of Figure 5a,b, respectively. Zero-shear viscosities $\eta_0$ are estimated from the plateau values at the low shear rate range. Figure 5a,b show the plot of $\eta_0$ as functions of salt concentrations of NaCl and KBr, respectively. The salt concentration dependence of $\varphi_p$ is presented in the insets of Figure 5a,b for samples with NaCl and KBr, respectively. We find that the Equation (4) can be used to fit the zero-shear viscosity data in Figure 5a,b using the values of $P$ obtained from the prediction of $a_s$ in the insets of Figure 4a,b for NaCl and KBr salts, thereby confirming the validity of the sticky-Rouse model and TTSS. Moreover, the value of $P$ used to fit the shift factors and zero-shear viscosities is in qualitative agreement, within order of magnitude, with the theoretically predicted value $P = \sqrt{200 N_A} e^3/2\pi(\varepsilon_0 \varepsilon k_B T)^{\frac{3}{2}} = 4.41$ calculated at $T_r$ = 20 °C and $C_{NaCl}^r$ = 4.6 M. For this calculation, we use $\varepsilon = 42$, which is the dielectric constant of a salt solution at the reference concentration of NaCl [44]. However, the effective dielectric constant of background medium could be even smaller due to the presence of high concentration of polyelectrolytes and get a closer agreement between experimental and theoretical values of $P$. Nonetheless, the current comparisons clearly show that the dynamics of the segments between two sticky points follows similar relaxation dynamics and can be described using the sticky-Rouse model across the precipitate-coacervate continuum. However, the observed gradual departure from the TTSS at lower frequencies and lower salt concentrations may suggest that the system may no longer be considered homogeneous and that longer collective relaxation times are present in the precipitate-coacervate continuum. These dynamics further slow down at longer length scales resulting in a rubbery relaxation as shown in Figure 4a,b.

## 3. Conclusions

We find that all polyelectrolyte complexes prepared at different salt concentrations exhibit TTS. TTS provides master curves that probe the relaxation behavior across the precipitate-coacervate continuum over wide time scales not previously investigated. The complexes behave liquid-like exhibiting a rheological transition from near-terminal relaxation to solid-like behavior at high frequencies and high salt concentrations. As the salt concentration is decreased systematically, a second transition associated with rubbery relaxation emerges at low frequencies and intermediate salt concentrations. The complexes show solid-like behavior as the salt concentration is reduced further. Moreover, we find that the complexes exhibit TTSS over the entire range of frequencies only at high salt concentrations. The width of this overlapped-frequency range decreases and the overlap shifts to higher frequencies as the salt concentration is reduced. This behavior may result from correlated collective dynamics of longer segments between ion pairs or perhaps entanglements at low salt concentrations and lower frequency (longer time scales). These TTSS trends are captured by the main results of the sticky-Rouse model in terms of the salt concentration dependence of the activation energy of the reversible breaking and reformation of ion pairs across the precipitate-coacervate continuum.

## 4. Materials and Methods §

*4.1. Materials*

Sodium poly(styrene sulfonate) (NaPSS) of mass-average relative molar mass $M_w$ = 200,000 g/mol and poly(diallyl dimethyl ammonium chloride) (PDADMAC) of $M_w$ = 150,000 g/mol were purchased from Sigma–Aldrich, Inc (St. Louis, MO,USA). The degree of polymerization is approximately 970 when estimated from the ratio of $M_w$ to monomer molecular mass. Polyelectrolyte complexes were prepared using two different salts, either NaCl (Sigma–Aldrich, Inc.) or KBr (Sigma–Aldrich, Inc.). The first set of samples used NaPSS and PDADMAC with NaCl. The second set used KBr with potassium-poly(styrene sulfonate) (KPSS) and poly(diallyl dimethyl ammonium bromide) (PDADMAB).

The second set of polymers with alternate counterions were prepared by ion-exchange. The solution of PDADMAB is obtained in a single step ion-exchange process by running 2% PDADMAC solution through $Br^-$ resin (Br-exchanged DOWEX MR-3, Sigma–Aldrich, Inc.). On the other hand, the solution of KPSS is obtained from 2% NaPSS in a two-step process. In the first step, $Na^+$ ions of NaPSS is replaced by $H^+$ using $H^+$ ion-exchange resin, Dowex 650C (Sigma–Aldrich, Inc. St. Louis, MO, USA). Subsequently, the $H^+$-exchanged polyelectrolytes are converted to KPSS by titration using KOH at pH = 7. The solutions of PDADMAB and KPSS are finally freeze-dried to obtain dry powders.

Polyelectrolyte complexes were prepared by mixing aqueous solutions of anionic and cationic polymers at 1:1 stochiometric ratio of the ion-containing groups. An initial concentration of 0.15 mol/L of monomers is used for all the polyelectrolyte solutions. Predetermined salt concentrations in the solutions of anionic and cationic polyelectrolyte are adjusted separately before mixing. The two solutions of oppositely charged polyelectrolyte are next mixed using gravimetric vortex for 30 s. The mixed-solution that undergoes phase separation is left undisturbed for three hours before performing centrifugation at $g = 10,000$ m/s$^2$ for 15 min. This results in polymer-poor supernatant at the top and polymer-rich complex at the bottom of microcuvette (Figure 1). Subsequently, all samples are annealed at 60 °C for 48 h after preparation and then stored at room temperature for 5 d before performing any measurements. Evaporation of water from the samples was prevented by sealing the cuvettes with high vacuum grease and parafilm wax.

*4.2. Rheological Characterization*

All rheological measurements are performed on Ares G2 stain-controlled rheometer using cone–plate geometry with a cone radius of 25 mm and a cone angle of 1°. The surfaces of the geometry are sand-blasted to a roughness value of approximately 2 μm to avoid slippage of the sample. After transferring the required amount of the coacervate from the bottom of the sample cuvette, a small amount of silicone oil of viscosity 4.6 mPa·s was applied to the rim of the cone fixture to avoid evaporation of water. No sign of degradation was observed in samples when in contact with the silicone oil for more than a week. Frequency sweep measurements are performed in the frequency range of 0.02 to 100 rad/s and at strain amplitude of 0.5 % to 5 % depending on the torque response of various samples. The applied strain is much smaller compared to the extent of the linear viscoelastic regime. A series of measurements is performed at nine temperature values in the range 10 to 60 °C. Samples were equilibrated at each temperature for at least 30 min before performing measurements. The time-temperature superposition and time-temperature-salt superposition of the rheological data are performed using the algorithm included in the TRIOS v4 software of TA Instruments (New Castle, DE, USA). To determine the zero-shear viscosity at different temperatures and salt concentrations, flow-curve (viscosity vs. shear rate) measurements are performed in the shear rate range $0.8 \times 10^{-5}$ to $0.1$ s$^{-1}$. However, the lower limit of the shear rate is adjusted depending on the viscosity of samples to maintain the measured torques above the noise-level of the instrument. The zero-shear viscosities were further determined from the intercept of the viscosity-plateau at low shear rate regime.

Polymer concentrations in complexes at different salt concentrations are estimated by measuring the volume of supernatant using a micropipette of volume resolution of 1 μL. However, the process of separating supernatant from the complex leads to a maximum uncertainty of 20 μL, which is within 2% of the total sample volume. Finally, the polymer concentration in complexes is calculated from the known values of volume ratio of supernatant-complex phases and initial polymer concentration assuming that the amount of polymer in supernatant phase is negligible compared to the complex, which is often a valid assumption for a 1:1 charge-stochiometric mixture [7].

§ Certain commercial equipment and materials are identified in this paper in order to specify adequately the experimental procedure. In no case does such identification imply recommendation by the National Institute of Standards and Technology (NIST) nor does it imply that the material or equipment identified is necessarily the best available for this purpose.

**Acknowledgments:** This work was supported by the National Institute of Standards and Technology (NIST) Materials Genome Initiative. We are grateful to Gale Holmes (NIST) for access to the rheology facilities used in this study, Miguel Vega (Texas Tech University) and Thomas Wu (Northwestern University) for their help with the ion-exchange and sample preparation processes during the NIST Summer Undergraduate Research Fellowship program, and Debra Audus (NIST) for fruitful discussions and critical comments.

**Author Contributions:** Samim Ali and Vivek M. Prabhu conceived and designed the experiments; Samim Ali performed the experiments; Samim Ali and Vivek M. Prabhu analyzed and wrote the paper.

**Conflicts of Interest:** The authors declare no conflict of interest.